# Monte Carlo Modeling of Spin FETs Controlled by Spin-Orbit Interaction

Min Shen,[a] Semion Saikin,[a,b,c] Ming-C. Cheng[a] and Vladimir Privman[a,b]

[a] Center for Quantum Device Technology,
[a] Department of Electrical and Computer Engineering, and
[b] Department of Physics,
Clarkson University, Potsdam, New York 13699-5720, USA

[c] Department of Physics, Kazan State University,
Kazan, Russian Federation

## Abstract

A method for Monte Carlo simulation of 2D spin-polarized electron transport in III-V semiconductor heterojunction FETs is presented. In the simulation, the dynamics of the electrons in coordinate and momentum space is treated semiclassically. The density matrix description of the spin is incorporated in the Monte Carlo method to account for the spin polarization dynamics. The spin-orbit interaction in the spin FET leads to both coherent evolution and dephasing of the electron spin polarization. Spin-independent scattering mechanisms, including optical phonons, acoustic phonons and ionized impurities, are implemented in the simulation. The electric field is determined self-consistently from the charge distribution resulting from the electron motion. Description of the Monte Carlo scheme is given and simulation results are reported for temperatures in the range 77-300 K.

# 1. Introduction

The Monte Carlo approach has been a widely used scheme for simulation of submicron or deep-submicron semiconductor devices. With given material properties of the semiconductor, it can account for non-equilibrium phenomena of charge carrier transport in the device channel and provide resolution beyond the drift-diffusion and hydrodynamic models. The step-wise simulation feature of the Monte Carlo approach makes it easier to incorporate different physics in the simulation [1] and avoids the assumptions needed in deriving alternative continuum drift-diffusion and hydrodynamic models [2,3]. It is because of this advantage that Monte Carlo simulation can also be used to provide the physical parameters required as the input data for drift-diffusion and hydrodynamic models.

During the recent years, spin-polarized electron transport in semiconductors became an active research topic due to its promise of applications in novel devices [4-7]. Many devices utilizing spin-dependent phenomena have been proposed [8-17]. The basic idea is to use the additional spin degree of freedom, which is usually ignored in charge-transport models, to encode information in the spin-polarized current. The design of new spintronic devices requires control for the spin polarization in the device channel. Recent experimental advances [5] have allowed efficient injection of spin-polarized current into low-dimensional semiconductor structures [18,19] and its maintenance for up to few nanoseconds at room temperature [20]. Generally, the electron spin dynamics can be controlled by external magnetic field, local magnetic fields produced by magnetic impurities and nuclei, and spin-orbit interaction. These interactions lead to coherent evolution of carrier spin polarization and also cause spin dephasing.

There are different approaches to describe the spin-polarized current in different transport regimes. Quantum-mechanical single-particle models have been utilized for ballistic spin-polarized electron transport [15,21,22]. Semiclassical drift-diffusion models have been derived based on the two-current (spin-up and spin-down) approximation [23-25] or the full spin-polarization vector description [26]. Recently, some nonlinear corrections in spin-polarized electron transport have attracted attention [27]. Boltzmann equations for two spin states [28] and for the spin density matrix [29,30] have also been considered. The Monte Carlo simulation approach has been applied for investigation of the spin polarized transport properties in semiconductor 1D and 2D structures in the presence of a moderate electric field [31-34]. The simulation results are promising and consistent with the existing experimental data.

Using the Monte Carlo method with incorporation of the spin density matrix dynamics [34], the present study reports new simulations for spin-polarized electron transport in an FET channel modeled as a single quantum well of a III-V heterostructure. Although the methodology [34] is simple and involves certain assumptions on the device structure, it is expected to be applicable beyond the regime of the drift-diffusion transport model. Moreover, additional details of the structure can be easily incorporated within this approach. During the simulations, the Poisson equation is solved for every sampling time step to update the electric field in the device channel. Electrons injected from the source have random momentum directions and the Maxwellian distribution of magnitudes which is related to the lattice temperature of the semiconductor. Both isotropic and anisotropic scattering processes are considered.

In Section 2, we review the model and describe the implementation of the Monte Carlo procedure. In Section 3, we apply the Monte Carlo method to study the spin-polarized dynamics in a representative device channel structure for a spin-FET. Results for the spin-polarization vector and the temperature dependence of the spin dephasing length in a spin FET subjected to different applied voltages are presented and discussed in Sections 3 and 4. Section 5 is devoted to a short summarizing discussion.

## 2. Implementation of the Spin Density Matrix Dynamics in Monte Carlo Simulation

We start with the Hamiltonian of a single conduction electron, including its spin,

$$H(\boldsymbol{\sigma},\mathbf{k}) = H_0(\mathbf{k}) \cdot \hat{1}_s + H_s(\boldsymbol{\sigma},\mathbf{k}) \ . \tag{1}$$

Assuming that the external magnetic field is zero, $\hat{1}_s$ on the right-hand side of Eq. (1) is the unity operator in the spin variables; $H_0$ is the spin-independent self-consistent single-electron Hamiltonian in the Hartree approximation,

$$H_0 = -\frac{\hbar^2}{2m^*}k^2 + V_H(\mathbf{r}) + H_{e\text{-}ph} + H_{ph} + V_{imp} \ , \tag{2}$$

The term $V_{imp}$ describes ionized nonmagnetic impurities, quantum well roughness and other static imperfections of its structure. The terms labeled "e-ph" and "ph" represent the electron-phonon interaction and the phonon mode Hamiltonian, respectively. The Hartree potential $V_H$ accounts for the electron-electron interactions. It is determined by the appropriate Poisson equation [35],

$$\nabla^2 V_H = -\frac{e^2}{\varepsilon_s}\left(\sum_j |\psi_j(\mathbf{r})|^2 - N_D\right), \quad (3)$$

where $\varepsilon_s$ is the material permittivity, $|\psi_j(\mathbf{r})|^2$ is the probability density to find the $j^{th}$ electron at $\mathbf{r}$, and $N_D$ is the ionized donor concentration. The second term on the right-hand side of Eq. (1) describes the spin dependent interactions with magnetic impurities and nuclear spins, and also the spin-orbit interaction. In this work, we only consider the effects of the spin-orbit interaction, which has been identified [36] as the main cause of spin relaxation in III-V semiconductors at high temperatures, 77-300 K.

An appropriate description of the electron spin in an open quantum system can be given by the spin density matrix [37],

$$\rho_\sigma(t) = \begin{pmatrix} \rho_{\uparrow\uparrow}(t) & \rho_{\uparrow\downarrow}(t) \\ \rho_{\downarrow\uparrow}(t) & \rho_{\downarrow\downarrow}(t) \end{pmatrix}, \quad (4)$$

where $\rho_{\uparrow\uparrow}$ and $\rho_{\downarrow\downarrow}$, both real numbers in [0,1], add up to 1 and represent the probabilities to find the electron with spin up or spin down. The off-diagnal matrix elements $\rho_{\uparrow\downarrow}$ and $\rho_{\downarrow\uparrow}$, which are complex-conjugate of each other, describe the degree of superposition of the spin-up and spin-down states. The density matrix (4) can be parameterized by the three (real-number) electron spin-polarization vector components, defined as $S_\zeta(t) = Tr(\sigma_\zeta \rho_\sigma(t))$, where $\zeta = x, y, z$, and $\sigma_\zeta$ are the Pauli matrices [37].

To specify the spin-orbit interaction terms, we consider a single III-V asymmetric quantum well grown in the (0, 0, 1) crystallographic direction in a spin FET. The main spin-orbit contributions in this case arise due to the Dresselhaus mechanism [38,39],

$$H_D = \beta \langle k_z^2 \rangle (k_y \sigma_y - k_x \sigma_x), \quad (5)$$

namely the bulk inversion asymmetry of the crystal, and Rashba mechanism [40],

$$H_R = \eta(k_y\sigma_x - k_x\sigma_y) \ ,  \qquad (6)$$

caused by the inversion asymmetry of the quantum well. To specify the momentum and spin-polarization vector components, we use the coordinate system where *x* is the direction of the electric field along the channel, while *z* is orthogonal to the quantum well plane. Moreover, the axes are oriented along the principal crystal axes, and the quantum well is assumed to be narrow, such that $k_x^2, k_y^2 \ll \langle k_z^2 \rangle$. The latter properties are important for the assumed form of the Dresselhaus spin-orbit interaction term in Eq. (5) [39].

For submicron or deep-submicron devices with smooth potential, in the considered temperature regime (*T* = 77-300 K), the spatial electron dynamics can be assumed semiclassical and described by the Boltzmann equation [35]. The electrons travel along classical "localized" trajectories between the scattering events. The scattering rates are given by the Fermi's golden rule, and the scattering events are instantaneous [35]. The phonon bath in Eq. (2) is assumed to remain in thermal equilibrium with the constant lattice temperature *T*. In this case, the Monte Carlo approach can be applied to the spatial transport [1-3]. We assume here that the influence of the electron spin evolution on the spatial motion is negligible owing to the small value of the electron momentum-state splitting due to the spin-orbit interactions in comparison with its average momentum. This is consistent with the original model of the D'yakonov-Perel' spin-relaxation mechanism [41].

In the simulation model, electrons propagate with constant momentum during the time interval which is the smaller of the time left to the next sampling time *t* + Δ*t* and the time left to the next scattering event (see the two "free flight calculation" blocks in the flowchart in Fig. 1(a)). The propagation momentum is set equal to the average value of the momentum of a particle moving with constant acceleration during this time interval. We term this motion "free flight." For each "free flight" time interval, $\tau$, the spin density matrix evolves according to

$$\rho_\sigma(t+\tau) = e^{-i(H_R+H_D)\tau/\hbar} \rho_\sigma(\tau) e^{i(H_R+H_D)\tau/\hbar} \ . \qquad (7)$$

Equation (7) is equivalent to rotation of the spin polarization vector about the effective magnetic field determined by the direction of the electron momentum. We assume that there is no electron spin-flip event accompanying momentum scattering [42]. The exponential operators in Eq. (7) can be written as 2 × 2 scattering matrices,

$$e^{-i(H_R+H_D)\tau/\hbar} = \begin{pmatrix} \cos(|\alpha|\tau) & i\frac{\alpha}{|\alpha|}\sin(|\alpha|\tau) \\ i\frac{\alpha^*}{|\alpha|}\sin(|\alpha|\tau) & \cos(|\alpha|\tau) \end{pmatrix}, \quad (8)$$

with the Hermitean conjugate of Eq. (8) for the operator $e^{i(H_R+H_D)\tau/\hbar}$. The appropriate sampling time step $\Delta t$ should be short as compared to all the dynamical time scales, in the Monte Carlo simulation. In Eq. (8), $\alpha$ is determined by the spin-orbit interaction terms given in Eqs. (5) and (6),

$$\alpha = \hbar^{-1}\left[\left(\eta k_y - \beta\langle k_z^2\rangle k_x\right) + i\left(\eta k_x - \beta\langle k_z^2\rangle k_y\right)\right]. \quad (9)$$

During the "free flight," the spin dynamics of a single electron spin is coherent; see Eq. (7). However, stochastic momentum fluctuations due to electron scattering events produce the distribution of spin states, thus causing effective dephasing at times $t > 0$.

The spin polarization, $\langle S_\zeta(\mathbf{r},t)\rangle$, of the current can be obtained by averaging $S_\zeta$ over all the electrons in a small volume $dv$, which is located at the space position $\mathbf{r}$, at time $t$. The absolute value of the average spin polarization vector is in the range $|\langle\mathbf{S}(\mathbf{r},t)\rangle| \leq 1$. If $|\langle\mathbf{S}(\mathbf{r},t)\rangle|$ is equal to 1, the electric current is completely spin-polarized. The components $\langle S_\zeta(\mathbf{r},t)\rangle$ define the orientation of the spin polarization, and evolution of the spin polarization vector may be viewed as consisting of coherent motion (rotation) and loss of polarization (reduction of magnitude) due to electron spin dephasing [39,41].

The implementation flowcharts of the Monte Carlo simulation approach are shown in Figs. 1(a) and 1(b). The flowchart in Fig. 1(a) is the subroutine that performs the single-particle simulation for the time interval between $t$ and $t+\Delta t$. The simulation given in this flowchart is carried out by sequentially performing spin rotation, free-flight and scattering calculations for one particle and its spin, if the time for the next-scattering-event time $t_s$ is less than $t + \Delta t$. After each scattering, the next-scattering-event time is updated as $t_s = t_s + \delta t_{scat}$, where $\delta t_{scat} = -(\ln p)/\Gamma$, and $p$ is a random number between 0 and 1, while $\Gamma$ is the total scattering rate including the self-scattering rate [2,3,43] that accounts for fictitious scattering introduced to make Γ constant. The sampling time step $\Delta t$ is specified small enough to properly update the particle motion and the electric field. The choice of the value of $\Delta t$ is based on the stability criteria [44]. The momentum increment and the distance of the "free flight" are calculated as

$$\Delta \mathbf{k} \hbar = -e\mathbf{E}\tau \ , \ \Delta \mathbf{r} = \frac{\hbar \ (\mathbf{k} + \Delta \mathbf{k}/2)}{m}\tau , \qquad (10)$$

where $e$ is the electron charge and $\mathbf{E}$ is the applied electric field. Based on the above discussion, the additional calculation needed to follow the spin polarization evolution of each particle, consists of an update of the spin density matrix at the beginning of each "free flight" time step, by using Eqs. (7) and (8).

It is assumed that the electrons are confined in the 1$^{st}$ (lowest) subband and that their $z$-direction motion is steady-state and defined by the shape of the quantum well. In the scattering event calculations, three in-plane ($xy$) scattering mechanisms are included in the simulation: optical phonon scattering, acoustic phonon scattering (for the scattering rates, see Sec. 2.6 of [44]), and separated impurity scattering (for the scattering rate, see Sec. 7 of [45]). The selection of the scattering mechanism is performed by defining

$$\Lambda_n(E_\mathbf{k}) = \sum_{j=1}^{n} W_j(E_\mathbf{k})/\Gamma, \quad n = 1, \ 2, \ 3 , \qquad (11)$$

where $W_j(E_\mathbf{k})$ is the integral scattering rate for the $j^{th}$ mechanism. The $n^{th}$ scattering mechanism is chosen if a random number $p$ falls between $\Lambda_{n-1}(E_\mathbf{k})$ and $\Lambda_n(E_\mathbf{k})$. In the scattering calculation, the in-plane projection of the electron momentum $k' = |\mathbf{k}'|$ is obtained from the energy conservation relation as $k' = \sqrt{2mE_{\mathbf{k}'}}/\hbar$, where $E_{\mathbf{k}'} = E_\mathbf{k} \pm \hbar\omega$ for the optical phonon scattering, and $E_{\mathbf{k}'} = E_\mathbf{k}$ for the acoustic-phonon and impurity scattering.

The flowchart shown in Fig. 1(b) is the main program for the Monte Carlo simulation which implements the simulation of many particles by repeatedly calling the subroutine for one particle simulation shown in Fig. 1(a). It also specifies the initial states of the particles, enforces the boundary conditions and updates the charge distribution and the self-consistent electric field in the channel. The following boundary conditions are assumed. Electrons are injected at the emission boundary with the kinetic energy

$$E = -k_B T \ln p \tag{12}$$

($T$ is the lattice temperature), and the injection angle (with respect to the $x$ axis) is randomly distributed between $-\pi/2$ and $\pi/2$. The electrons that fly beyond the collection boundary (and some that return through the injection boundary) are absorbed, and a new electron is emitted whenever there is an electron absorbed. This process ensures that the total number of the electrons in the device is constant in the simulation. The electric potential is the solution of the Poisson equation with the boundary conditions specified by the voltage applied to the device.

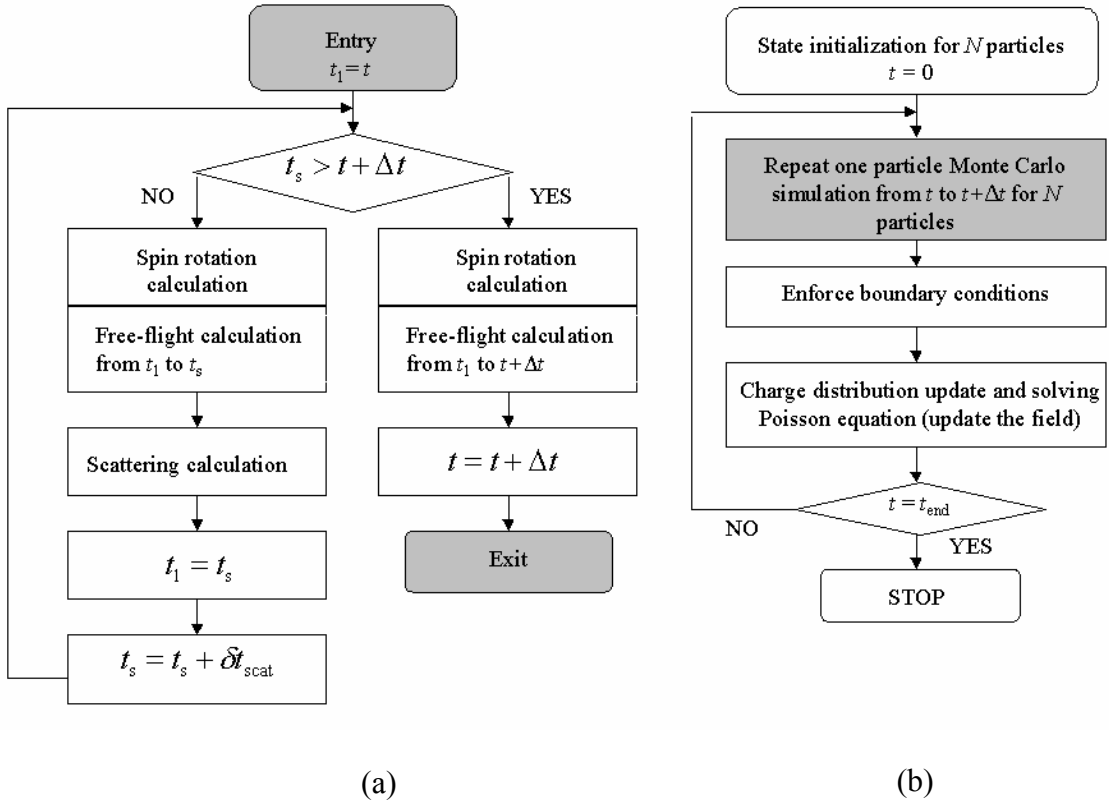

                (a)                 (b)

**Fig. 1.** Flowcharts for the Monte Carlo simulation: (a) the single-particle Monte Carlo calculation, including the spin density matrix update, from $t$ to $t + \Delta t$; and (b) the ensemble Monte Carlo calculation for $N$ particles.

## 3. Simulation results.

For simulations, we have used the structure with the 0.55 μm channel length and infinite width, Fig. 2(a). The confining potential is assumed to be that of a single asymmetric $In_{0.52}Al_{0.48}As/In_{0.53}Ga_{0.47}As/In_{0.52}Al_{0.48}As$ quantum well, Fig. 2(b), in the one-subband approximation. The width of quantum well is $d = 20$ nm. The structure is n-doped with donor concentration $N_D = 10^{12}$ cm$^{-2}$. We assume that all the donors are ionized, and the equilibrium electron concentration in the channel is equal to $N_D$. The calculated energy of the 1$^{st}$ subband is $E_1 \approx 0.2$ eV. The energy splitting between the 1$^{st}$ and 2$^{nd}$, excited, subband is estimated as $\Delta E_{12} \approx 60 - 70$ meV. This value in turn defines the range of the drain-source voltage values, $V_{DS}$, for which the one-subband approximation model is valid. The Rashba electron spin-orbit coupling constant used in the simulation was $\eta = 0.074$ eV·Å [46], while the value of the Dresselhaus constant, $\beta = 32.2$ eV·Å$^3$, was taken close to this parameter in bulk GaAs [47]. The material band structure and scattering parameters were adopted from [48]. In the simulation, the total number of particles in the channel was $N = 55000$, and the sampling time step was $\Delta t = 1$ fsec. To achieve the steady-state transport regime, we ran the simulation program for 20000 time steps, and collected data only during the last 2000 time steps.

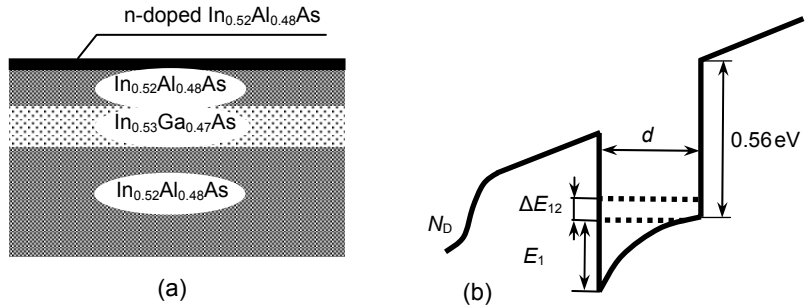

**Fig. 2.** (a) The device structure, and (b) the confining potential.

The simulated in-channel electron concentration and energy profile are shown in Fig. 3. For the considered range of the applied voltages, the steady-state charge distribution in the device channel is nearly constant except for the source boundary. Owing to the utilized boundary conditions, the charge accumulation layer is generated near the injection boundary, Fig. 3(a). The injection region, which is about 0.02μm, can be considered quasi-ballistic, where electrons experience strong acceleration, Fig. 3(b). Because of the low applied voltage, the transport in the rest of the device is effectively drift-diffusive.

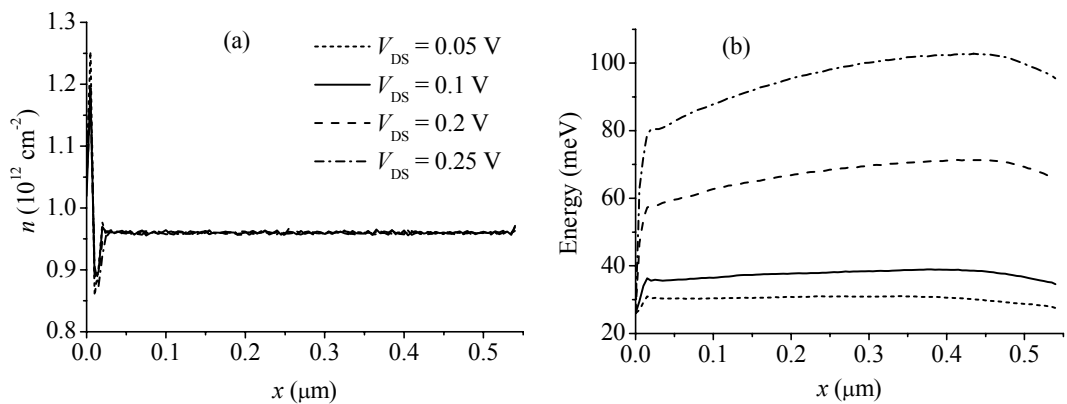

**Fig. 3.** The calculated electron transport parameters: (a) electron concentration in the channel, and (b) average energy profile, as functions of $x$, at $T = 300$ K and $V_{DS} =$ 0.05-0.25 V.

The simulated steady-state distributions of the spin polarization for three different injected polarizations: along the $x$, $y$, and $z$ axes, are shown in Fig. 4. Due to anisotropy of the spin-orbit interaction terms in Eqs. (5) and (6), the spin dephasing rate is different for different orientations of the spin polarization in the drift-diffusive transport region. This leads to modulation of spin dephasing as a function of $x$ for the spin-polarized current with the injected spin polarization along the $x$ and $z$ directions, Fig. 4(d). For these cases, the spin polarization vector largely rotates in the $xz$-plane, Figs. 4(a) and 4(c). The dephasing will be stronger for the polarization vector oriented in the $z$ direction. This can be explained as follows. In the considered structure, the Rashba spin-orbit coupling is considerably stronger than the Dresselhaus coupling, $\eta/(\beta\langle k_z^2\rangle) \approx 5.3$. Thus, the term proportional to $k_y$, see Eq. (6), is primarily responsible for the spin dephasing [8]. It will not affect the polarization vector oriented in the $x$ direction, due to proportionality to $\sigma_x$. This effect is responsible for the variation of the spin relaxation time with the orientation of the injected spin polarization discussed in [33].

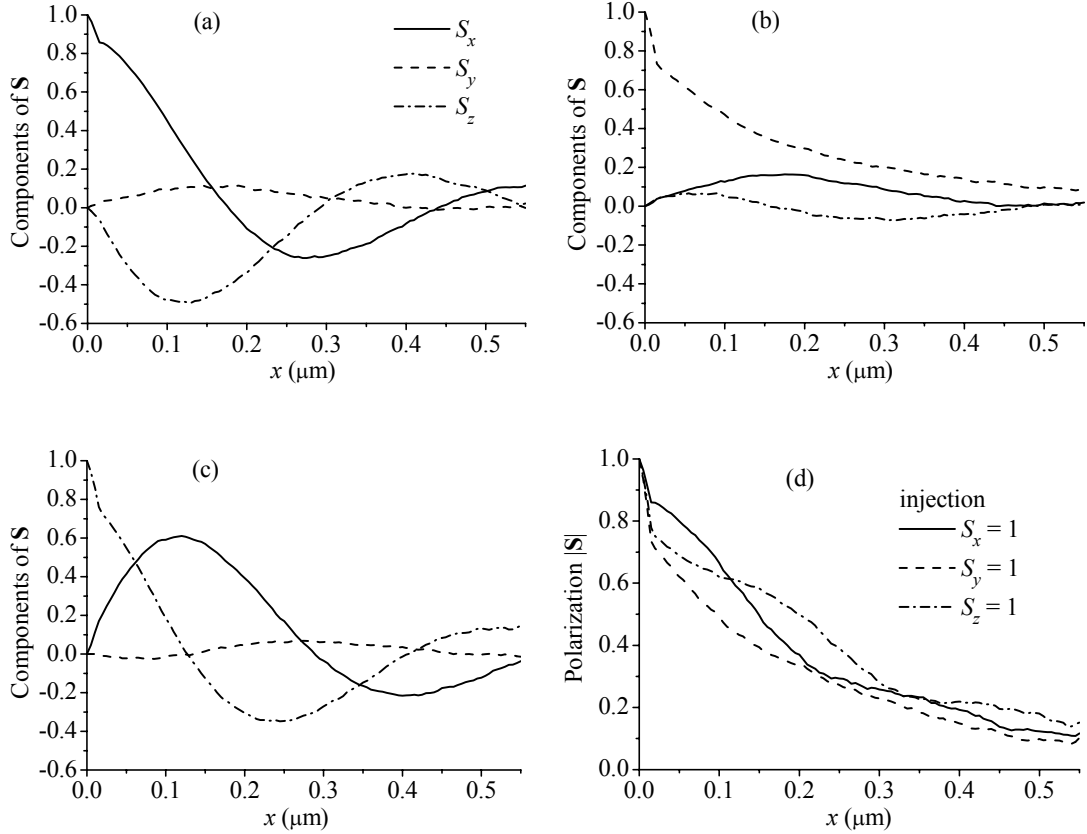

**Fig. 4.** The steady-state spin polarization, **S**, in the channel, for $V_{DS} = 0.1$ V, $T = 300$ K, for three different injected polarizations. (a)-(c) The components of the spin polarization vector. (d) The magnitude of the spin polarization vector.

Similar to coherent oscillations of the spin polarization in the FET structure, Fig. 2, the spin dephasing is also affected by the electron transport properties. In the quasi-ballistic transport region, the spin polarization decreases significantly, as can be seen in Fig. 4(d). This can be affected by the electron velocity distribution. In Figs. 5 and 6, we present the correlation between the ratio of the electron thermal energy to the drift energy and the electron spin dephasing for different applied voltages and temperatures, respectively. Both figures show a common characteristic that the spin dephasing is strongly dependent on this energy ratio. It can be seen from the figures that high ratio corresponds to faster dephasing, while low ratio corresponds to slow dephasing. This is consistent with the expectation that the more random is the electron motion in space, the more efficient will the spin dephasing be. The effect of the injected electron energy on the drop of the initial spin polarization was also obtained in the simulation [49] of spin polarized transport in bulk GaAs.

Because the injected electrons are randomly distributed in **k** space in the $+x$ direction, the thermal energy is considerably greater than the drift energy near $x = 0$. As a result, the ratio of the thermal to drift energy is large, and fast spin dephasing is observed. Due to small energy (thus weak scattering) near $x = 0$, electrons are strongly accelerated over a small distance ($l \sim 0.01$-$0.02$ μm) influenced by the rapidly increasing electric field before undergoing strong scattering. Over this small distance, the electron average velocity grows faster than the average energy, and the ratio is drastically reduced, as shown in Figs. 5 and 6. The minimum of the ratio near $x = 0.01$ μm is actually caused by the velocity overshoot [34]. At higher voltage, the drift velocity is larger, which results in the smaller ratio and slower spin dephasing, as displayed in Fig. 5. At higher temperature, the thermal energy is greater, which gives rise to the larger ratio and faster spin dephasing, as shown in Fig. 6.

Although the velocity overshoot at $x \sim 0.01$ μm reduces the ratio significantly, fast spin dephasing is still observed up to $x \sim 0.015$ μm. This might be an artifact of the insufficient spatial resolution of the numerical simulation. It should be noted that there are only four mesh points in $0 < x < 0.02$ μm. The detailed correlation between the energy ratio and dephasing may not be captured exactly with such space resolution.

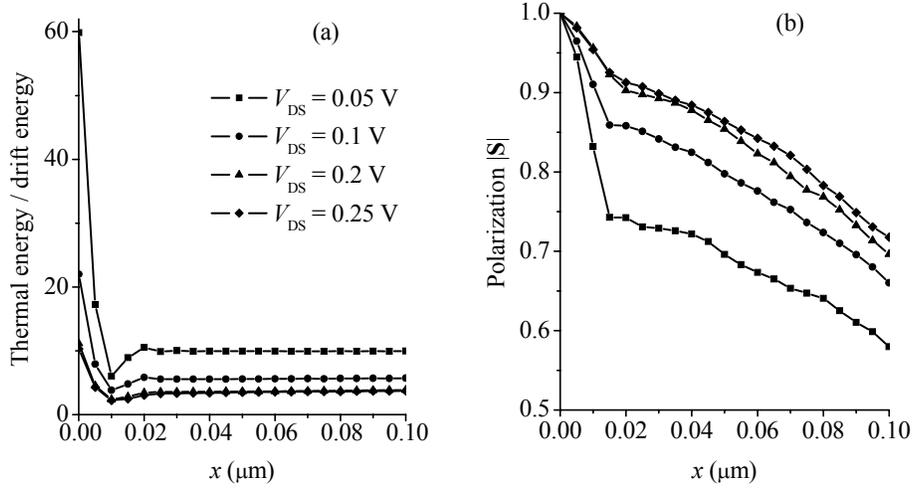

**Fig. 5.** The observed correlation between the ratio of the thermal to drift energy and spin dephasing for different applied voltages, at $T$ = 300 K. (a) The ratio of electron thermal energy to drift energy, and (b) spin dephasing.

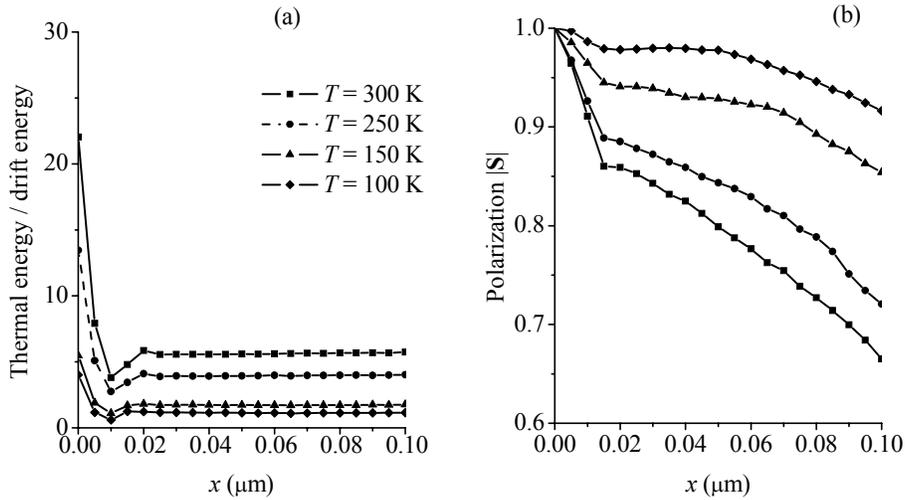

**Fig. 6.** The observed correlation between the ratio of the thermal and drift energies and spin dephasing for different temperatures, at $V_{DS}$ = 0.1V. (a) The ratio of electron thermal energy to drift energy, and (b) spin dephasing.

The spin dephasing along the channel is not a simple exponential decay. However, we identify the spin dephasing or spin scattering length, $l_s$, as the distance over which the spin polarization is reduced by the factor of $e$ from the injected value. This parameter can be used for rough estimation of a length scale within which spin-dependent phenomena are important for a given structure. In Fig. 7, we show the spin dephasing length for injected spin polarization parallel to the *x* axis for different temperatures and applied voltages. For higher values of $V_{DS}$, at low temperatures spin depolarizes faster, Fig. 7. This can be attributed to the effect of stronger scattering. However, at room temperature we observe the opposite dependence, due to larger drop of polarization in the ballistic region for smaller applied voltage. For constant drain-source voltage, the spin dephasing length is almost linearly dependent on the temperature in the considered range.

## 4. Discussions

Our simulation model has incorporated the leading D'yakonov-Perel'-type spin dephasing mechanism only, which should be dominant in the semiclassical transport regime. For more accurate estimations of the electron spin dephasing, additional mechanisms should be considered [50].

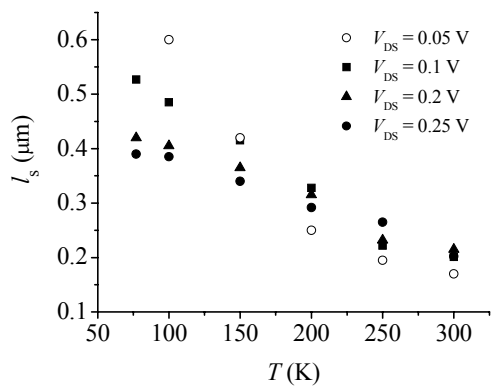

**Fig. 7.** Spin dephasing length as a function of the temperature for different values of the applied voltage (for the injected spin polarization $S_x = 1$).

In narrow band gap semiconductors such as InGaAs, the Elliott-Yafet spin-dephasing mechanism [42] can play an important role. Due to admixing of the hole states in the conduction electron wave functions, the electron spin can flip with some probability even at a non-magnetic impurity. This mechanism can be integrated in the Monte Carlo scheme in the scattering calculation, together with the momentum scattering. Another possible spin dephasing mechanism arises due to the electron-electron interaction [29,51].

The validity of the one-subband approximation model is in doubt for room-temperature electron transport. In the considered case, it can be argued that the inter-subband electron scattering only contributes corrections to spin dephasing [34]. However, for more accurate calculations, inter-subband processes should be incorporated into the simulation model.

The specific device structure can also lead to additional spin dephasing mechanisms. For example, the current spin dephasing due to magnetic field created by the ferromagnetic source and drain in a spin-FET [52] may be more critical than the considered D'aykonov-Perel'-type spin relaxation.

## 5. Conclusions

A Monte Carlo method for simulation of spin-polarized electron transport in submicron spin-FET structures has been developed. The electron spin polarization is described by the spin density matrix, while the spatial dynamics of the electron is treated semiclassically. The coherent dynamics of the current spin polarization and spin dephasing are determined by the spin-orbit interaction. The electric field in the device is evaluated self-consistently with the charge distribution. The phonon and impurity electron momentum scattering mechanisms are incorporated in the simulation. The steady state spatial distribution of the current spin-polarization vector has been simulated. The temperature dependence of the spin dephasing length was calculated for the range of 77-300K. The estimated value of the spin dephasing length at room temperature is of the order of 0.2 μm in the $In_{0.52}Al_{0.48}As/In_{0.53}Ga_{0.47}As/In_{0.52}Al_{0.48}As$ FET structure with the quantum well grown in the (0, 0, 1) crystallographic direction.

*Acknowledgement* - We thank Drs. L. Fedichkin, A. Shik and I. D. Vagner for helpful discussions. This research was supported by the National Security Agency and Advanced Research and Development Activity under Army Research Office contract DAAD-19-02-1-0035, and by the National Science Foundation under grant DMR-0121146.

**References.**